\pdfoutput=1
\documentclass[journal,12pt,onecolumn,draftclsnofoot]{IEEEtran}
\IEEEoverridecommandlockouts
\usepackage{cite}
\usepackage{subfigure}
\usepackage[ruled,vlined,algo2e]{algorithm2e}
\usepackage{algorithm}
\usepackage{algorithmic}
\usepackage{graphicx}
\usepackage{amssymb}
\usepackage{amsmath}
\usepackage[numbers, sort]{natbib}
\begin{document}
\title{Semantic Communication for Cooperative Perception using HARQ}

\author{Yucheng Sheng,~\IEEEmembership{Graduate Student Member,~IEEE,}
        Le Liang,~\IEEEmembership{Member,~IEEE,}
        \\Hao Ye,~\IEEEmembership{Member,~IEEE,}
        Shi Jin,~\IEEEmembership{Fellow,~IEEE,}
        and Geoffrey Ye Li,~\IEEEmembership{Fellow,~IEEE}
%\thanks{Corresponding author: Le Liang}
\thanks{Yucheng Sheng, and Shi Jin are with the National Mobile Communications Research Laboratory, Southeast University, Nanjing 210096, China (e-mail: shengyucheng@seu.edu.cn; jinshi@seu.edu.cn).}
\thanks{Le Liang is with the National Mobile Communications Research Laboratory, Southeast University, Nanjing 210096, China, and also with Purple Mountain Laboratories, Nanjing 211111, China (e-mail: lliang@seu.edu.cn).}
\thanks{Hao Ye is with the Department of Electrical and Computer Engineering, University of California, Santa Cruz, CA 95064, USA (e-mail: yehao@ucsc.edu).}
\thanks{Geoffrey Ye Li is with the ITP Lab, the Department of Electrical and Electronic Engineering, Imperial College London, SW7 2BX London, U.K. (e-mail: geoffrey.li@imperial.ac.uk).}
}

\maketitle

\begin{abstract}
Cooperative perception, offering a wider field of view than standalone perception, is becoming increasingly crucial in autonomous driving. This perception is enabled through vehicle-to-vehicle (V2V) communication, allowing connected automated vehicles (CAVs) to exchange sensor data, such as light detection and ranging (LiDAR) point clouds, thereby enhancing the collective understanding of the environment. In this paper, we leverage an importance map to distill critical semantic information, introducing a cooperative perception semantic communication framework that employs intermediate fusion. To counter the challenges posed by time-varying multipath fading, our approach incorporates the use of orthogonal frequency-division multiplexing (OFDM) along with channel estimation and equalization strategies. Furthermore, recognizing the necessity for reliable transmission, especially in the low SNR scenarios, we introduce a novel semantic error detection method that is integrated with our semantic communication framework in the spirit of hybrid automatic repeated request (HARQ). Simulation results show that our model surpasses the traditional separate source-channel coding methods in perception performance, both with and without HARQ. Additionally, in terms of throughput, our proposed HARQ schemes demonstrate superior efficiency to the conventional coding approaches.
\end{abstract}




\IEEEpeerreviewmaketitle

% Main text

\section{Introduction}
Shannon's separation theorem in the 1940s has long been a cornerstone in the field of communication. It suggests that source and channel coding could be optimized separately to approach ideal transmission rates with sufficiently large coding blocks \cite{shannon}. However, modern wireless communication, driven by the urgency of real-time applications such as the Internet of Things (IoT) and autonomous vehicles, necessitates a departure from this traditional paradigm. The infinite block lengths, idealized in Shannon's theory, clash with the practical demands for low latency and minimal computational complexity in time varying wireless communication scenarios. To address these challenges, semantic communication with joint source-channel coding (JSCC) has been proposed. This method tailors the coding process to specific tasks, optimizing both source and channel coding in unison for improved performance. The initial exploration of JSCC, applied to text transmission, utilized recurrent neural networks (RNNs) to improve semantic fidelity \cite{farsad2018deep}, \cite{xie2021}. Concurrently, deep learning-based JSCC systems were designed to directly convert image pixel values into complex channel input symbols, bypassing the traditional separate stages of compression and channel coding \cite{jscc-image}, \cite{DeepJSCC-f}, \cite{DeepJSCC-Q}. Moreover, in the domain of video transmission, JSCC strategies have been applied to video compression, channel coding, and modulation into a single neural network-based operation, streamlining the transmission process and optimizing bandwidth usage \cite{DeepWiVe}, \cite{Video}.

Essential for autonomous vehicles, sensor data, particularly light detection and ranging (LiDAR) point clouds, contains critical semantic content \cite{xu2022opv2v}. Autonomous vehicles often face limited perception capabilities due to obstructions, such as buildings and trees. To circumvent these limitations, the concept of cooperative perception has been explored, where connected automated vehicles (CAVs) share their sensor data to achieve a composite view of their surroundings \cite{xu2022opv2v}, \cite{zhou2022multi}, \cite{arnold2020cooperative}, \cite{tu2022maxim}, \cite{li2023learning}, \cite{hu2022where2comm}. Vehicle-to-vehicle (V2V) communication is the cornerstone of this system, facilitating the exchange of a wide array of data types, from raw sensory inputs to processed detection outputs, along with crucial metadata, such as timestamps and positional information. Cooperative perception leverages various fusion techniques, which can be categorized into early, intermediate, and late fusion, each with distinct communication demands. Early fusion, which transits raw LiDAR point clouds, demands the highest amount of resources. Intermediate fusion processes and condenses raw data into more manageable features, striking a balance between resource usage and performance efficacy. Late fusion focuses on sharing the final detection results from each CAV, conserving resources but potentially sacrificing accuracy. Typically, intermediate fusion is preferred in cooperative perception due to its efficiency in resource utilization while retaining the potential to achieve the same level of performance as early fusion. 

Building on intermediate fusion, various innovative solutions have been investigated to strike a balance between perception performance and communication overhead. For instance, the selective handshake mechanism in \cite{liu2020when2com} identifies and connects the most pertinent CAVs. This mechanism employs a learning-based approach to dynamically form communication groups and determines the optimal timing for data exchange. In addition, the innovative approach to source coding through an end-to-end learning framework \cite{xu2022v2x} employs a spatially aware graph neural network (GNN) to effectively aggregate information collected from CAVs. However, these methods often assumes flawless communication between CAVs, disregarding the potential channel impairments. To address this gap, the novel semantic communication scheme for cooperative perception in \cite{sheng2023semantic} showcases the potential of semantic communication in enhancing cooperative perception.

Due to safety considerations in autonomous driving, data reliability becomes particularly crucial in cooperative perception. Previous approaches cannot guarantee the transmission reliability under low signal-to-noise ratio (SNR) conditions, posing a significant challenge for the deployment of these technologies in real-world scenarios. Traditionally, hybrid automatic repeat request (HARQ) protocols have been utilized in communication systems to bolster transmission reliability. However, there is limited research on the application of HARQ within semantic communication systems, especially those customized for cooperative perception. Existing studies \cite{Video}, \cite{jiang2022deep} typically center around the design of error detection methods and aim to ensure the integrity of the reconstructed data rather than the successful completion of task-oriented objectives. These methods, while valuable for reconstruction tasks, may not fully align with the specific requirements of applications, such as cooperative perception, where the ultimate goal extends beyond mere data reconstruction. Designing error detection methods and HARQ protocols for task-oriented semantic communication systems, such as cooperative perception, is challenging. It encompasses not only ensuring the robustness and reliability of data transmission in adverse communication environments but also aligning the error correction mechanisms with the semantic goals of the system.

In response, we first propose a novel semantic error detection method to identify semantic mistakes in the task-oriented semantic communication system. Additionally, we present a cooperative perception semantic communication framework that integrates HARQ, thereby enhancing the reliability of data transmission. Our system is built upon a JSCC architecture and is developed through end-to-end learning, which is optimized to achieve better semantic performance with HARQ. Our main contributions can be summarized as follows:
\begin{itemize}
	\item We introduce a model for cooperative perception based on JSCC that leverages an importance map to transmit semantic information effectively, aiming to enhance semantic performance while minimizing communication overhead. By conducting evaluations over a time-varying multipath fading channel, our proposed JSCC communication framework demonstrates superior performance compared to conventional methods that treat source and channel coding separately.
        \item  In contrast to traditional bit-error detection methods that rely on cyclic redundancy check (CRC), we develop SimCRC, a semantic error detector designed to assess whether received features necessitate retransmission. This detector employs a Siamese Network architecture and undergoes training via similarity ranking to accurately identify semantic errors.     
        \item To ensure the reliability of our model under diverse channel conditions, we have integrated our JSCC framework with HARQ techniques. This integration consists of a chase combining strategy, designated as SimHARQ-I, and an approach that incorporates additional redundancy, termed SimHARQ-II. We have conducted extensive simulation on time-varying multipath fading channels to show the proposed SimHARQ-I and SimHARQ-II schemes outpeform conventional separate source and channel coding approaches in both perception performance and throughput.

\end{itemize}
The remainder of this paper is structured as follows: Section II delineates our proposed JSCC model tailored for cooperative perception. Section III elaborates on our novel semantic error detection approach, outlines the HARQ designs, and details the training methodologies employed. Section IV presents empirical evidence showcasing the advantages of our proposed model through simulation results. Finally, Section V concludes the paper.

\section{System Model and Problem Formulation}
\begin{figure*}[htp]
    \centering
    \includegraphics[width=16cm]{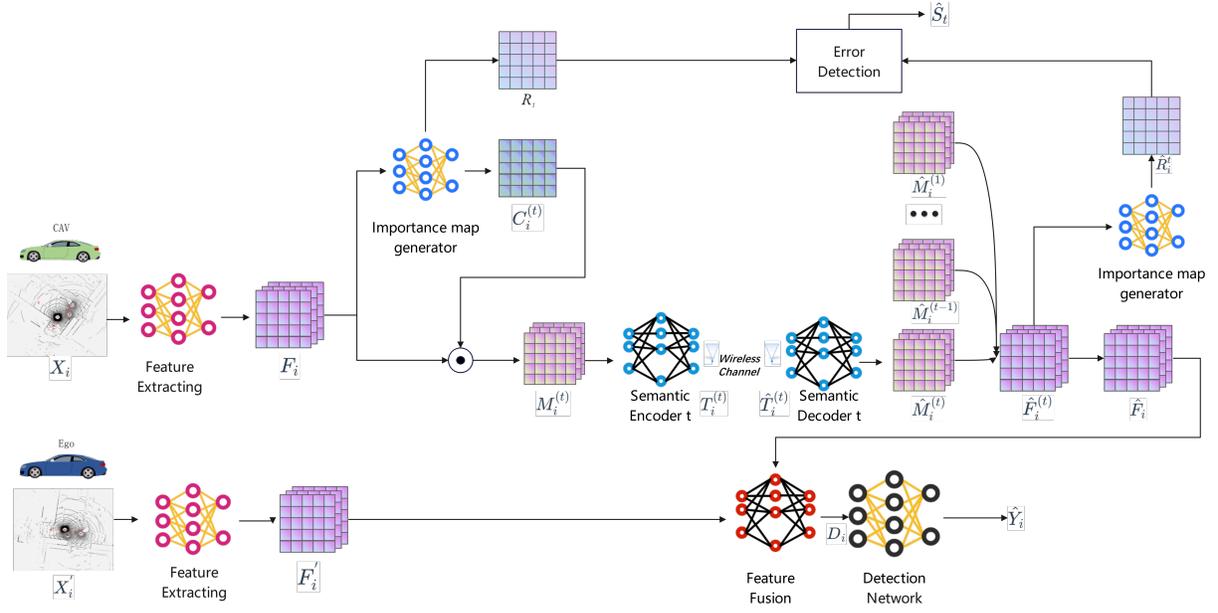}
    \caption{Structure of the cooperative perception model based on the importance map at the $t$th transmission.}
    \label{fig:framework2}
\end{figure*}

Consider a scenario where one CAV shares intermediate features with an ego car. To strike a balance between perception accuracy and communication overhead, we employ an importance map generator to identify the most crucial elements within the entire feature set. The cooperative perception system with intermediate fusion is structured around four key components: feature extraction, feature sharing, feature fusion, and detection result generation. In this section, we first present the overall procedure of cooperative perception based on the importance map. Subsequently, we extend our cooperative perception model to an OFDM-based framework, aiming to address the challenges posed by time-varying multipath fading channels. Lastly, we will discuss the intricacies and challenges associated with implementing HARQ within the cooperative perception framework.

\subsection{Cooperative Perception Model based on Importance Map}

The framework depicted in Fig. \ref{fig:framework2} outlines cooperative perception  with intermediate fusion at the $t$th transmission, where the anchor-based PointPillar method \cite{lang2019pointpillars}, \cite{chen2019cooper}, \cite{zhou2018voxelnet}, \cite{liu2020when2com} is chosen as the backbone to extract the semantic information of the $i$th sample from the raw LiDAR point clouds $X_i$. The extracted feature tensor, $F_i$, in the CAV can be represented as
\begin{equation}
    F_i=\Phi(X_i),
\end{equation}
where $F_i\in{\mathbb{R}^{C\times{H}\times{W}}}$, $C$ is the number of channels, $H$ and $W$ represent the height and width of the feature tensor, respectively. The extracted feature tensor $F^{'}_i$ for the ego car can be defined similarly.

In the context of feature sharing, the configurations of the semantic encoder and decoder may vary, depending on the specific HARQ scheme, which will be elaborated in the subsequent section. Here, we introduce a versatile framework for feature sharing, elucidating the process through which feature tensors are sent to the ego vehicle over multiple transmissions. As depicted in Fig. \ref{fig:framework2}, the importance map generator, $P(\cdot)$, is tasked with pinpointing critical elements within the feature tensor, subsequently generating an importance map $C_i^{(t)}$ for the $t$th transmission. Following this, the data for the $t$th transmission, $M_i^{(t)}$, is generated via element-wise multiplication of $F_i$ with the importance map $C_i^{(t)}$, denoted by
\begin{equation}    
    C_i^{(t)}=P(F_i),
\end{equation}
and
\begin{equation}
    M_i^{(t)}={F_i}\odot{C_i^{(t)}}, 
\end{equation}
respectively, where $C_i^{(t)}\in{[0,1]^{{H}\times{W}}}$ and $M_i^{(t)}\in{\mathbb{R}^{C\times{H}\times{W}}}$. In this study, we define the compression ratio, $\text{CR}^{(t)}=\frac{M_i^{(t)}}{F_i}$, as the ratio of the data size of the transmitted feature tensor $M_i^{(t)}$ at the $t$th transmission to the size of the entire feature tensor $F_i$. Through extensive experimentation, we find that setting $\text{CR}^{(t)}$ to the order of $10^{-2}$ achieves a desirable balance between conserving transmission resources and maintaining perception accuracy. Given the sparsity of $M_i^{(t)}$,  the overall transmission requirement is substantially reduced.

Then we develop a semantic encoder $\Psi_s^{(t)}(\cdot)$ and a semantic decoder $\Psi_d^{(t)}(\cdot)$ specifically for the $t$th transmission.  Different from the traditional communication frameworks, our system integrates source coding, channel coding, and modulation within a unified scheme, which will be trained in an end-to-end manner, to enhance semantic-level transmission efficacy. Particularly, the complex symbol stream for the $t$th transmission is derived from $M_i^{(t)}$ using the semantic encoder, denoted by
\begin{equation}
    T_i^{(t)}=\Psi_s^{(t)}(M_i^{(t)}),
\end{equation}
where $T_i^{(t)}\in{\mathbb{C}^{H^{'}\times{W^{'}}}}$, $H^{'}$ and $W^{'}$ represent height and width of the tensor $T_i^{(t)}$. Following the encoding process, the joint source-channel coded sequence, $T_i^{(t)}$, is sent over a time-varying multipath fading channel. 

At the receiver side (the ego car), the received symbol, denoted as $\hat{T}_i^{(t)}$, is mapped to semantic information by the semantic decoder $\Psi_d^{(t)}(\cdot)$, yielding $\hat{M}_i^{(t)}$,
\begin{equation}
    \hat{M}_i^{(t)}=\Psi_d^{(t)}(\hat{T}_i^{(t)}).
\end{equation}

Considering that prior transmissions may contain correlated information, the candidate feature $\hat{F}_i^{(t)}$ for the $t$th transmission is formulated by aggregating all preceding messages up to the $t$th transmission. This aggregation is achieved through a function $f_c(\cdot)$, represented as
\begin{equation}
\hat{F}_i^{(t)} = f_c(\hat{M}_i^{(1)}, \ldots, \hat{M}_i^{(t-1)}, \hat{M}_i^{(t)}). \label{eq:fc}
\end{equation}

Upon passing error detection, $\hat{F}_i^{(t)}$ is updated to $\hat{F}_i$ and is utilized in the ensuing fusion steps. In scenarios where transmission attempts to reach their maximum, denoted by $B$, the selection of $\hat{F}_i$ from the collection ${\hat{F}_i^{(1)}, \hat{F}_i^{(2)}, \ldots, \hat{F}_i^{(R)}}$ is determined through a specific selection strategy, details of which will be discussed in the subsequent section.

Utilizing attention-based mechanisms for feature fusion, we integrate the semantic data, $\hat{F}_i^{(t)}$, with $F^{'}_i$, deriving a comprehensive fusion tensor that encapsulates semantic insights from both the CAV and the ego car. This fusion output, labeled as $D_i$, is formulated as

\begin{equation}
    D_i^{(t)} = \chi(\hat{F}_i, F^{'}_i), 
\end{equation}
where $D_i\in \mathbb{R}^{C \times H \times W}$ represents the fused tensor and $\chi(\cdot)$ symbolizes the attention fusion network. Notably, our method adopts a self-attention fusion approach, which is instrumental in elucidating the spatial interplay between $\hat{F}_i$ and $F^{'}_i$, enhancing the efficacy of the fusion process.

Following the feature fusion, a bounding box method can be generated through the prediction header along with the associated confidence scores. The detection output, $\hat{Y}_i$, generated by the detection network $\Gamma(\cdot)$, can be expressed as
\begin{equation}
    \hat{Y}_i=\Gamma(D_i),
\end{equation}
where $\hat{Y}_i$ consists of the regression output and classification output. The regression component includes the 3D position, dimensions, and yaw angle of the predefined anchor boxes. The classification component assigns a confidence score to each bounding box, indicating the probability that it encloses an object.
% The regression component includes seven key parameters: $(x, y, z, w, l, h, \theta)$, which detail the 3D position, dimensions, and yaw angle of the predefined anchor boxes. These parameters are essential for adjusting the anchor boxes to align accurately with the objects detected. On the other hand, the classification component assigns a confidence score to each bounding box, indicating the probability that it encloses an object. The combination of regression and classification results yields the final predictions from the object detection framework, playing a vital role in accurate object identification and spatial localization.

\subsection{Cooperative Perception Model with OFDM}
For each feature transmission in OFDM systems, a distinct time slot is allocated, comprising $N_p$ pilot symbols for channel estimation and $N_s$ data symbols for payload transmission. Channel estimation is facilitated using block-type pilots, where known symbols are transmitted across all subcarriers over a series of OFDM symbols. The complex symbol stream, $T_i$, for the $t$th transmission is subject to an initial normalization step, resulting in $T_i^n \in{\mathbb{C}^{H^{'}\times{W^{'}}}}$, represented as
\begin{equation}
T_i^n = \sqrt{H^{'}W^{'}P}\frac{T_i}{\sqrt{T_i^{*}T_i}},
\end{equation}
where $(\cdot)^{*}$ is the conjugate transpose and $P$ represents the average transmit power constraint. The normalized symbol stream, $T_i^n$, is reshaped into $T_i^m \in \mathbb{C}^{N_s \times L_{\text{fft}}}$, where $L_{\text{fft}}$ denotes the number of subcarriers in an OFDM symbol.

If the product of $H'$ and $W'$ is less than that of $N_s$ and $L_{\text{fft}}$, zero-padding is applied to reshape $T_i^n$ into $T_i^m$. The pilot symbols, denoted as $T_i^p \in \mathbb{C}^{N_p \times L_{\text{fft}}}$, are crucial for channel estimation. In our approach, the $N_p$ pilot symbols are synthesized using Quadrature Amplitude Modulation (QAM) on randomly generated bits. Once the pilot and data symbols are prepared, we employ the Inverse Fast Fourier Transform (IFFT) to convert from frequency to time domain, resulting in $T_i^p$ and $T_i^m$. This step is followed by appending a cyclic prefix (CP) to enhance the resilience to multipath effects by reducing intersymbol interference. The composite transmit signal, $E$, encompassing both pilot and data symbols post-CP addition, is then represented as $E \in \mathbb{C}^{(N_p + N_s) \times (L_{\text{fft}} + L_{\text{cp}})}$, ready for transmission over the channel. At the receiver, upon acquiring the noisy channel output, $\hat{E}$, the system first discards the CP and then employs the Fast Fourier Transform (FFT) to revert to the frequency domain, yielding the received pilot symbols, $\hat{T}_i^p$ and data symbols, $\hat{T}_i^m$. 

In this paper, we consider wireless communication over a slowly time-varying multipath fading channel through an OFDM system without intercarrier interference. We assume that the Doppler spread is significantly smaller than the subcarrier spacing, allowing the channel to be considered static throughout the duration of an OFDM symbol. Thus, the received frequency domain symbols of the pilots and information can be represented as
\begin{equation}
    \hat{T}_i^m[j,k]=H[j,k]T_i^m[j,k]+Z[j,k],
\end{equation}
and
\begin{equation}
    \hat{T}_i^p[j,k]=H[j,k]T_i^p[j,k]+Z[j,k],
\end{equation}
where $H[j,k]$ is the channel frequency response at the $k$th subcarrier of the $j$th OFDM symbol, and $Z[j,k]$ denotes the AWGN. Meanwhile, $H[j,k]$ can be represented as
\begin{equation}
    H[j,k]=\sum_{m=0}^{M-1} a_{m}(j)e^{-j2\pi k\Delta f \tau_m},
\end{equation}
where $M$ is the number of taps, $\Delta f$ is the subcarrier spacing, and $(a_{m}(j),\tau_m)$ represent the amplitude and delay of the $m$th channel tap. It is noteworthy that $a_{m}(j)$ is a time-dependent function, varying with time indice $j$. We adopt the minimum mean-squared error (MMSE) approach for channel estimation and employ a conventional MMSE equalizer for equalization.

% We adopt the minimum mean-squared error (MMSE) approach for channel estimation, expressed as
% \begin{equation}
% {H}_{\text{MMSE}}[j] = R_{HH_{\text{LS}}}\left(R_{HH} + \frac{1}{\text{SNR}}I\right)^{-1}H_{\text{LS}},
% \end{equation}
% where $R_{HH_{LS}}$ is the cross-correlation matrix between the actual channel response $H$ and the least squares (LS) channel estimation $H_{\text{LS}}$ for the $j$th pilot symbol. $R_{HH}$ is the autocorrelation matrix of $H$, SNR is the signal-to-noise ratio, and $I$ is the identity matrix of the same dimensions as $R_{HH}$. In simulations, the real channel response $H$ is approximated by $H_{\text{LS}}$ when computing $R_{HH}$ and $R_{HH_{\text{LS}}}$, as the exact channel response is typically unknown at the receiver. For equalization, we employ a conventional MMSE equalizer formulated as
% \begin{equation}
% \hat{T}_i[j,k] = \frac{H_{\text{MMSE}}[j,k]^{*}\hat{T}_i^m[j,k]}{|H_{\text{MMSE}}[j,k]|^2 + \sigma^2},
% \end{equation}
% where $H_{\text{MMSE}}[j,k]$ denotes the estimated channel response for the $j$th OFDM symbol and the $k$th subcarrier, and $\sigma^2$ is the noise power. This process aids in mitigating the effects of channel noise and distortion, improving the accuracy of the reconstructed source information.

\subsection{HARQ in Cooperative Perception}

Conventional communication systems utilize three main HARQ mechanisms, namely HARQ-I, HARQ-II, and HARQ-III, based on the types of retransmission content \cite{HARQsurvey}.

\textbf{HARQ-I: Chase Combining.} HARQ-I, also known as the traditional HARQ scheme, transmits the same data packet in all retransmissions. It involves adding cyclic redundancy check (CRC) bits to the transmitted data packets and applying forward error-correction (FEC) encoding. The receiver performs FEC decoding and CRC check on the received data. If errors are detected, the receiver will discard the data associated with the erroneous group and send a negative acknowledgment (NACK) signal, requesting the retransmission of the same data from the previous frame.

\textbf{HARQ-II: Incremental Redundancy.} HARQ-II selectively transmits parity bits as required, giving it the designation ``incremental redundancy". 
Hence, in HARQ-II, the transmitted data during the initial transmission and subsequent retransmissions typically differ. Unlike HARQ-I, the receiver does not discard the previously transmitted erroneous groups. Instead, it combines them with the received retransmitted groups for improved decoding. 

\textbf{HARQ-III.} Generally, HARQ-III, closely resembles HARQ-II with the distinction that both information data and parity bits are incorporated in each retransmission. Therefore, each data packet can be independently decoded or synthesized into a codeword with more significant redundancy for combined decoding.

Integrating the proposed semantic communication framework with traditional HARQ mechanisms faces new challenges. Unlike conventional systems where CRC is used for error detection, our proposed method transmits features directly, bypassing the need for quantization and CRC. This necessitates the development of a semantic-based error detection method to identify semantic errors as traditional CRC cannot be applied to unquantized data. Moreover, to design novel HARQ mechanisms that align with JSCC in our system, similar to traditional HARQ-I and HARQ-II strategies but with semantic considerations. It is an open area for research. We aim to adapt the principles of chase combining and incremental redundancy to suit our cooperative perception communication framework.

\section{Cooperative perception with HARQ}
In this section, we first introduce a novel semantic error detection method, named SimCRC. Subsequently, the training methodology of SimCRC is outlined. Then, we  delineate the architecture of our cooperative perception system that integrates HARQ-I, referred to as SimHARQ-I, as well as the architecture employing incremental redundancy, designated as SimHARQ-II. The loss function and corresponding training algorithm of our proposed method are also explicated in detail.

\subsection{Semantic Error Detection Method}
In traditional transmission systems, CRC error detection is typically employed to facilitate automatic retransmission requests, ensuring the correct reception of transmitted data alongside feedback acknowledgments (ACKs). However, the conventional CRC method might flag a transmission as erroneous if the bit-error rate (BER) is non-zero, even when the transmitted content is semantically very close to the original. In semantic communication, the focus shifts from bit-level precision to semantic similarity. For instance, in sentence semantic communication systems \cite{jiang2022deep}, the BERT model, a pre-trained neural network, is utilized to assess sentence similarities. However, many task-oriented semantic communication systems, including our proposed cooperative perception framework, do not inherently possess a mechanism like BERT for semantic similarity evaluation. Therefore, we propose a semantic-based error detection approach, termed SimCRC, and develop a novel method to measure semantic similarity between the original feature $F_i^{(t)}$ and its reconstructed counterpart $\hat{F}_i^{(t)}$, in terms of a similarity score $S$.
\begin{figure}[htp]
    \centering
    \includegraphics[width=13cm]{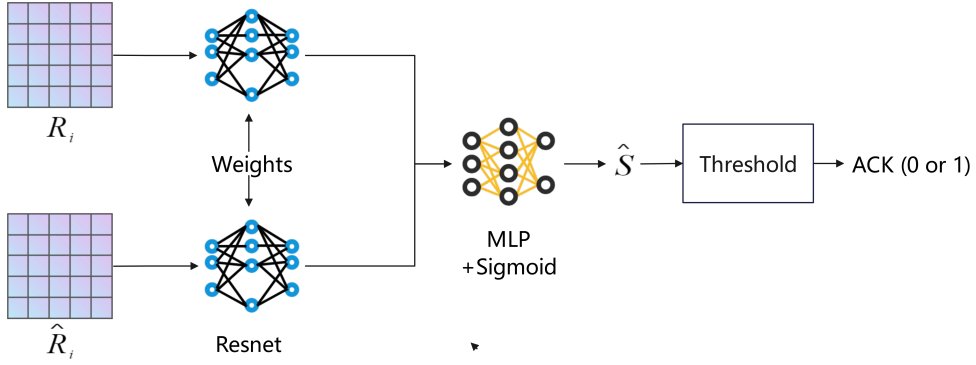}
    \caption{Structure of the similarity prediction based on the siamese neural network.}
    \label{fig:error detection}
\end{figure}

It is widely known that to some extent, the loss can represent the semantic distance between the predicted result and the ground truth. The perception loss, $L_{per}$, in cooperative perception is composed of localization loss $L_{local}$ and confidence loss $L_{conf}$, represented as
\begin{equation}
L_{per} = L_{local} + L_{conf},\label{eq:per}
\end{equation}
where $L_{local}$ quantifies the deviation in position, size, and yaw angle between predicted and actual anchor boxes, and $L_{conf}$ assesses the discrepancy between predicted confidence scores and true labels. Semantic errors in $\hat{F}_i^{(t)}$ will elevate $L_{per}$, indicating degraded perception performance.

To compute semantic similarity, we examine the difference between the perception losses of the original and reconstructed features. The similarity score $S$ is inversely related to this difference, modeled as
\begin{equation}
S = \min(U, -\log_{10}|L_{per}(F_i^{(t)}) - L_{per}(\hat{F}_i^{(t)})|),
\end{equation}
where $U$ is a predefined upper limit for $S$. This score helps determine semantic closeness between $F_i^{(t)}$ and $\hat{F}_i^{(t)}$. However, we need to train a similarity score $\hat{S}$ to predict the true score $S$, since $F_i^{(t)}$ is unavailable during operation.

Prior information, analogous to CRC in the traditional systems, is necessary for semantic error detection. Before sharing features, vehicles broadcast metadata like relative pose and confidence maps \cite{hu2022where2comm}. As depicted in Fig. \ref{fig:framework2}, confidence maps $R_i$ and $\hat{R}_i$, derived from the original and reconstructed features, contain vital semantic information and are smaller in size compared to the full feature, making them suitable as prior information. Note that \( R_i \)  and \( C_i^{(t)} \) are generated from different output layers of the importance map. \( C_i^{(t)} \) is a selection matrix composed of 0 and 1, whereas \( R_i \)  contains certain semantic information.  As Fig. \ref{fig:error detection} illustrates, by employing a siamese neural network \cite{chicco2021siamese} to feature dual resnet branches with shared weights, we generate the predicted semantic similarity $\hat{S}$ using $R_i$ and $\hat{R}_i$ as 
\begin{equation}
\hat{S} = \text{Sigmoid}(\text{MLP}([\text{Res}(R_i), \text{Res}(\hat{R}_i)])), \label{eq:hats}
\end{equation}
where $\hat{S}$ is then evaluated against a threshold $\beta$, a hyperparameter. If $\hat{S}$ is greater than $\beta$, ACK should be set as 1, confirming the semantic integrity of the received feature. If \(\hat{S}\) is less than \(\beta\), ACK should be set as 0, indicating significant semantic discrepancies.

\subsection{Training Algorithm of Semantic Error Detection Method}

While the predicted semantic similarity score $\hat{S}$ is intended to reflect the actual similarity score $S$, a precise numerical correspondence between the two is unnecessary. Consequently, using a regression loss function for the siamese network might not be the most effective approach. This ineffectiveness is attributed not only to the challenging nature of the training process but also to the susceptibility of the model to changes in the formulation of $S$. For instance, altering the function $f(x) = -\log_{10}(x)$ to $f(x) = -\log_{2}(x)$ could significantly impact the distribution of $S$, potentially necessitating a retraining of $\hat{S}$ without adding meaningful value.

Inspired by RankNet \cite{burges2010ranknet}, we propose to view it as a similarity ranking problem. Suppose the feature undergoes transmission under varying channel conditions $K$ times, yielding $K$ distinct pairs of $\hat{S}$ and $S$. We can easily justify that the feature from the $m$th transmission is semantically closer to the original than that from the $n$th if the similarity score $S_m$ from the $m$th transmission is higher than $S_n$ from the $n$th transmission. Consequently, we can establish a ranking of the $K$ transmissions based on their respective similarity scores $S$. The objective in RankNet training is to align the predicted scores, $\hat{S}$, with the ranking order defined by $S$. The training process aims to refine $\hat{S}$ so that it can accurately rank the transmission samples in a manner consistent with the ranking derived from $S$. This approach circumvents the need for a strict numerical match between $\hat{S}$ and $S$, focusing on preserving the relative ordering. It is more pertinent to semantic similarity assessment.

In the training phase, the dataset is segmented based on queries. It is important to note that the confidence map, $R_i$, remains unchanged across different samples. Consequently, the predicted score for the $m$th sample, as shown in  (\ref{eq:hats}), is reformulated as
\begin{equation}
\hat{S}_m = \text{Sigmoid}(\text{MLP}([\text{Res}(R_i), \text{Res}(\hat{R}_i^m)])), \label{eq:16}
\end{equation}
where $\hat{R}_i^m$ signifies the confidence map corresponding to the $m$th sampling.

For each query, the model examines every pair of samples, $S_m$ and $S_n$, along with their respective confidence map samplings $\hat{R}_i^m$ and $\hat{R}_i^n$. These pairs are processed through \ref{eq:16} to produce two outputs, which are then processed through a sigmoid function to estimate the probability that $\hat{S}_m$ is greater than $\hat{S}_n$, represented by
\begin{equation}
P_{mn} = P(\hat{S}_m > \hat{S}_n) = \frac{1}{1 + e^{-\boldsymbol{\sigma}(\hat{S}_m - \hat{S}_n)}},
\end{equation}
where $\boldsymbol{\sigma}$ is a hyperparameter that influences the curvature of the sigmoid function. The training objective is to minimize the discrepancy between the predicted probabilities and the actual ranking outcomes. The actual probability, denoted by $\overline{P}_{mn}$, reflects the ground truth that $\hat{S}_m$ is greater than $\hat{S}_n$. We can obtain \(\overline{P}_{mn}\) through $\overline{P}_{mn}=\frac{1}{2}(1+S_{mn})$, where $S_{mn}$ can be represented as
\begin{equation}
S_{mn}=\left\{
\begin{aligned}
1 & , & \text{if}\ S_m>S_n, \\
0 & , & \text{if}\ S_m=S_n,\\
-1 & , & \text{if}\ S_m<S_n.
\end{aligned}
\right.
\end{equation}

The cross entropy cost function, employed here, penalizes the deviation between the predicted $P_{mn}$ and the actual $\overline{P}_{mn}$, represented by
\begin{equation}
\begin{aligned}
 C_{mn} & = -\overline{P}_{mn} \log P_{mn} - (1 - \overline{P}_{mn}) \log(1 - P_{mn}) \\
        & = \frac{1}{2}(1-S_{mn})\boldsymbol{\sigma}(\hat{S}_m - \hat{S}_n)+\log{(1+e^{-\boldsymbol{\sigma}(\hat{S}_m - \hat{S}_n)})}. \label{eq:loss1}
\end{aligned}
\end{equation}

\begin{figure}[htp]
    \centering
    \includegraphics[width=5cm]{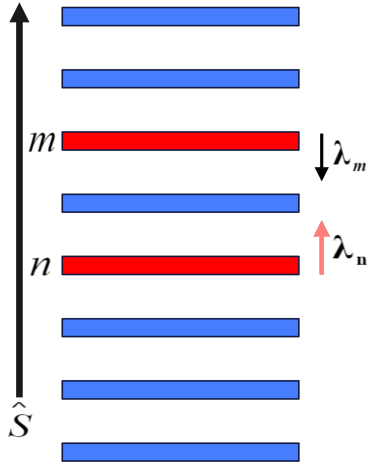}
    \caption{Visualization of the training process of RankNet \cite{burges2010ranknet}.}
    \label{fig:ranking}
\end{figure}

For a point cloud consisting of $K$ samples over different channels, there are $K(K-1)/2$ unique pairs of samples. While   (\ref{eq:loss1}) addresses the similarity ranking for individual pairs, the aggregate loss for all pairs is computed as 
\begin{equation}
C = \sum_{\{m,n\} \in I} C_{mn},
\end{equation}
where $I$ is the set comprising all unique index pairs $\{m, n\}$. To ensure each pair is included only once, we define that \( I \) contains pairs of indices \(\{m, n\}\) where \( S_m > S_n \), resulting in \( S_{mn} = 1 \). The gradient of the loss function with respect to a weight parameter $w_k$ is calculated as
\begin{equation}
\begin{aligned}
    \frac{\partial C}{\partial w_k}{\operatorname*{=}}\sum_{\{m,n\}\in I}\frac{\partial C_{mn}}{\partial w_k}{\operatorname*{=}}\sum_{\{m,n\}\in I}\lambda_{mn}\left(\frac{\partial \hat{S}_m}{\partial w_k}-\frac{\partial \hat{S}_n}{\partial w_k}\right) {\operatorname*{=}}\sum_m\lambda_m\frac{\partial \hat{S}_m}{\partial w_k},
\end{aligned}   
\end{equation}
where $\lambda_{mn}$ and $\lambda_m$ are weights that quantify the influence of each pair's gradient difference on the overall gradient. Specifically, $\lambda_{mn}$ is defined as
\begin{equation}
\lambda_{mn} = \sigma \left( \frac{1}{2}(1 - S_{mn}) - \frac{1}{1 + e^{\sigma(\hat{S}_m - \hat{S}_n)}} \right),
\end{equation}
which highlights the contribution of each pair to the gradient. The term, $\lambda_m$ is then the net effect of all pair-wise comparisons involving the $m$th sample, represented by
\begin{equation}
\lambda_m = \sum_{\{n | S_m>S_n, n \in I\}} \lambda_{mn} - \sum_{\{n | S_m<S_n, m \in I\}} \lambda_{mn}.
\end{equation}
Hence, \(\lambda_m\) can be regarded as the cumulative diminutive forces affixed to the $m$th sampling. As depicted in Fig. \ref{fig:ranking}, the orientation of these vectors indicates the cumulative direction and magnitude of adjustments needed for the predicted score $\hat{S}_m$ of $m$th sample. This approach ensures that the training process accounts for the relative rankings of all sample pairs, guiding the model to refine its predictions in a manner aligning with the overall similarity ranking structure of the dataset.

\subsection{Cooperative Perception System with HARQ}

\begin{figure*}[htp]
    \centering
    \includegraphics[width=16cm]{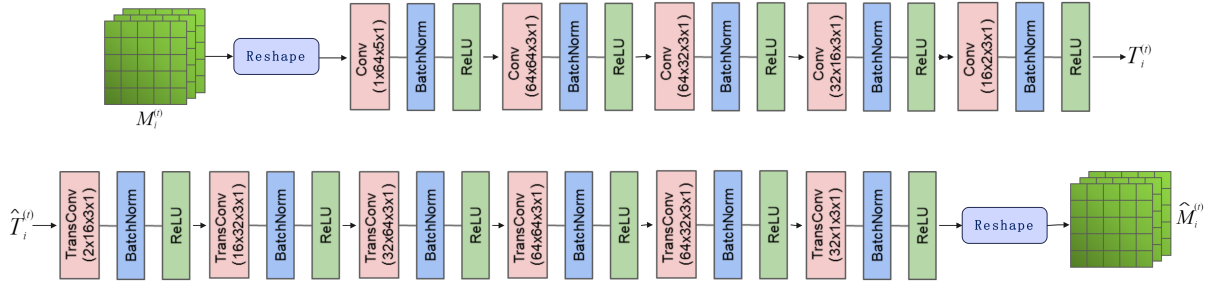}
    \caption{Network structure for our proposed semantic encoder and decoder at the $t$th transmission. The parameters for blue boxes are in the format of $ input ~channel ~size \times output~ channel ~size \times kernel~ size \times stride$.}
    \label{fig:network}
\end{figure*}

\begin{figure*}[htbp]
    \centering
    \begin{minipage}[t]{1\linewidth}
        \centering
        \includegraphics[width=0.7\textwidth]{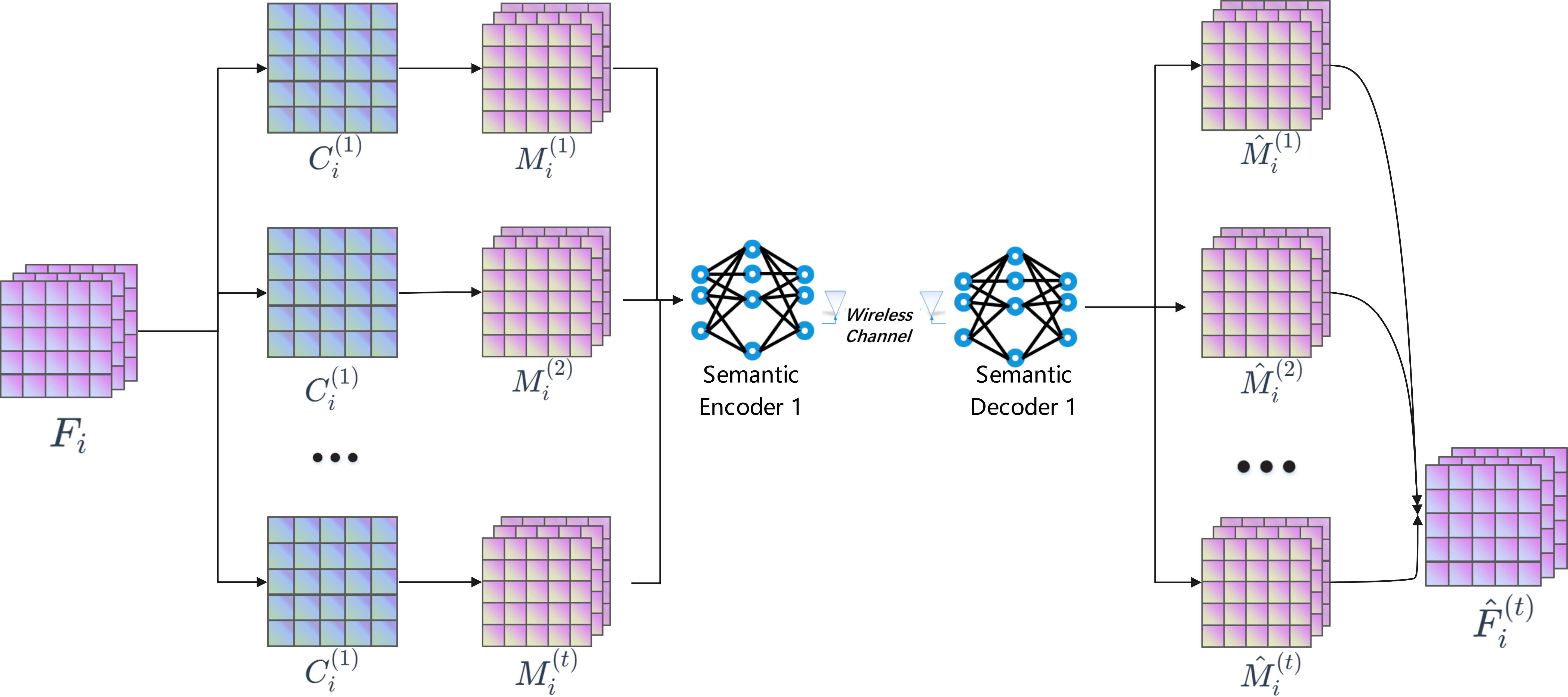}
        \centerline{(a)}
    \end{minipage}
    \vfill
    \begin{minipage}[t]{1\linewidth}
        \centering
        \includegraphics[width=0.7\textwidth]{HARQ-II.pdf}
        \centerline{(b)}
    \end{minipage}
    \caption{Structures of the cooperative perception with HARQ at the $t$th transmission. (a) SimHARQ-I; (b) SimHARQ-II.}
    \label{fig:HARQ12}
\end{figure*}

Initially, we present SimHARQ-I, an end-to-end semantic framework analogous to the HARQ-I framework. As illustrated in Fig. \ref{fig:HARQ12}(a), SimHARQ-I is engineered to transmit a feature repeatedly until the receiver successfully reconstructs it or the maximum transmission limit, denoted as $B$, is reached. Meanwhile, the feature can be decoded independently at the $t$th transmission. Considering that the transmitted information remains constant across various transmission rounds, it is possible that we use only one importance map, \(C_i^{(1)}\), and a unique pair of semantic encoder and decoder, \(\{\Psi_s^{(1)}, \Psi_d^{(1)}\}\).

We depict the architecture of the proposed semantic encoder and decoder in Fig. \ref{fig:network}. At the transmitter, the semantic encoder processes feature map \(M_i^{(1)}\) from \(\mathbb{R}^{C\times{H}\times{W}}\) to complex-valued channel input samples \(T_i^{(1)}\in\mathbb{C}^{H'\times{W'}}\). This process involves a deterministic function \(\Psi_s^{(1)}\) and converts \(\mathbb{R}^{C\times{H}\times{W}}\) to \(\mathbb{C}^{H'\times{W'}}\), adhering to an average power constraint. Initially, the feature map \(M\) is reshaped to eliminate the zero-component. The feature map is then transformed into \(T\) using a convolution neural network (CNN) composed of convolutional, normalization, and parametric ReLU activation layers. 

At the receiver, the semantic decoder converts received complex-valued samples \(\hat{T}_i^{(1)}\) from \(\mathbb{C}^{H'\times{W'}}\) back to the reconstructed feature map \(\hat{M}_i^{(1)}\in\mathbb{R}^{C\times{H}\times{W}}\) using a deterministic function \(\Psi_d^{(1)}\). The decoding process, employing transconvolutional layers, introduces greater complexity than the convolutional layers of the encoder. A reshape layer is then applied to align the reconstructed signal with \(\hat{M}_i^{(t)}\). Then, the $t$th candidate feature $\hat{F}_{i}^{(t)}$ can be obtained through (\ref{eq:fc}), where $f_c(\hat{M}_i^{(1)}, \ldots, \hat{M}_i^{(t-1)}, \hat{M}_i^{(t)})=\hat{M}_i^{(t)}$.

If the process reaches the transmission attempt limit, $B$, the selected feature \(\hat{F}_i\) is determined by
\begin{equation}
    \hat{F}_i= \hat{F}_{i}^{(t^*)},  \label{eq:fi}
\end{equation}
where  $t^* = \underset{t}{\mathrm{argmax}~\hat{S}_t}$ and \( \hat{F}_{i}^{(t^*)} \) is the feature from the transmission round \( t^* \) that maximizes the similarity score \( \hat{S}_t \), indicating that it closely resembles the original feature and is thus ideal for fusion. This approach ensures the selection of the most representative feature for subsequent processing stages.

Subsequently, we extend our investigation to an end-to-end semantic framework, SimHARQ-II, which is analogous to the HARQ-II framework. SimHARQ-II is devised to transmit incremental information until the feature is successfully reconstructed or the maximum number of transmission attempts is reached. Unlike SimHARQ-I, the whole transmitted revisions should be combined for decoding. To facilitate the $B$ transmission rounds, we employ two distinct pairs of semantic encoders and decoders, as depicted in Fig. \ref{fig:HARQ12}(b). The initial transmission utilizes the first encoder-decoder pair \(\{\Psi_s^{(1)}, \Psi_d^{(1)}\}\) while subsequent transmissions engage the second pair \(\{\Psi_s^{(2)}, \Psi_d^{(2)}\}\). The configuration of \(\{\Psi_s^{(1)}, \Psi_d^{(1)}\}\) is optimized for high SNR conditions, aiming to maximize efficiency and accuracy in favorable transmission environments. On the other hand, \(\{\Psi_s^{(2)}, \Psi_d^{(2)}\}\) is designed to be robust in lower SNR scenarios, ensuring reliable feature reconstruction under challenging conditions. 

For SimHARQ-II, the $t$th candidate feature $\hat{F}_{i}^{(t)}$ can be obtained through (\ref{eq:fc}), where \\$f_c(\hat{M}_i^{(1)}, \ldots, \hat{M}_i^{(t-1)}, \hat{M}_i^{(t)})=\hat{M}_i^{(1)}+\ldots+\hat{M}_i^{(t)}$. Moreover, the selection process of feature \( \hat{F_i} \) for fusion follows a methodology similar to that in SimHARQ-I, as specified in   (\ref{eq:fi}). This process ensures that the most appropriate feature maximizing a predefined similarity score is chosen for fusion, aligning with the overarching goal of achieving effective and accurate feature reconstruction across varying communication conditions.

\begin{algorithm2e}
\SetAlgoLined
\KwIn{The raw data $ X_i$  (LiDAR point clouds)}%输入参数
\KwOut{The detection output $\hat{Y}_i$}%输出
% \KwResult{Write here the result}
 \textbf{Step 1}: Train the semantic encoder and decoder. 
 \\
 \While{ not converge}{
   \For{$X_i$ in batch samples}{
         $F_i=\Phi(X_i)$\;
         $M_i^{(1)}={F_i}\odot{C_i}^{(1)}, C_i^{(1)}=P(F_i)$\;
         $T_i^{(1)}=\Psi_s^{(1)}(M_i^{(1)})$\;
         $\hat{M}_i^{(1)}=\Psi_d^{(1)}(\hat{T}_i^{(1)})$\;
   }
   Compute $L_{r}$\;
   Update the network $\Psi_s^{(1)}(\cdot)$,$\Psi_d^{(1)}(\cdot)$ using SGD;
   }
\textbf{Step 2}:  Train the whole network. 
 \\
 \While{ not converge}{
   \For{$X_i$ in batch samples}{
         $F_i=\Phi(X_i)$\;
         $F^{'}_i=\Phi(X^{'}_i)$\;
         $M_i^{(1)}={F_i}\odot{C_i}^{(1)}, C_i^{(1)}=P(F_i)$\;
         $T_i^{(1)}=\Psi_s^{(1)}(M_i^{(1)})$\;
         $\hat{M}_i^{(1)}=\Psi_d^{(1)}(\hat{T}_i^{(1)})$\;
         $D_i=\chi({\hat{F}_i,F^{'}_i})$\;
         $\hat{Y}_i=\Gamma(D_i)$\;
   }
   Compute $L_{total}$\;
   Update the whole network using SGD\;
   }
\caption{Training Algorithm without HARQ}
\end{algorithm2e}

% \begin{figure*}[htbp]
%     \begin{minipage}[t]{1\linewidth}
%         \centering
%         \includegraphics[width=0.5\textwidth]{HARQ-I.png}
%         \centerline{(a)}
%     \end{minipage}%
%     \begin{minipage}[t]{1\linewidth}
%         \centering
%         \includegraphics[width=0.5\textwidth]{HARQ-II.pdf}
%         \centerline{(b)}
%     \end{minipage}
%     \caption{Structures of the cooperative perception with HARQ at the $t$th transmission. (a) SimHARQ-I; (b) SimHARQ-II.}
%     \label{fig:HARQ12}
% \end{figure*}

\subsection{Loss function and Training Algorithm}
The training algorithm of our proposed system is structured in two distinct steps: the initial phase involves training without HARQ, followed by the training of SimHARQ-I and SimHARQ-II frameworks.

\subsubsection{Training Algorithm without HARQ}

By refining the reconstruction loss, the semantic encoder and decoder aim to diminish the average discrepancy between feature map \(M\) and its reconstructed counterpart \(\hat{M}\), as produced by the decoder. This reconstruction loss is quantified through the average mean-squared error (MSE) between \(M\) and \(\hat{M}\), expressed as
\begin{equation}
    L_{r}=\frac{1}{N}\sum_{i=1}^N ||M_i-\hat{M}_i||_2^2,
\end{equation}
where $N$ indicates the total number of samples. To assess the efficacy of cooperative perception, we implement  smooth loss for regression tasks and focal loss for classification tasks \cite{hu2022where2comm}, as indicated by
\begin{equation}
    L_{p}=\frac{1}{N}\sum_{i=1}^N L_{per}(Y_i,\hat{Y}_i),
\end{equation}
where \(L_{per}(\cdot)\) denotes the perception loss for one sample, as described in  (\ref{eq:per}). The overall loss function combines these elements, yielding
\begin{equation}
    L_{total}=\lambda L_{r} + L_{p},
\end{equation}
where \(\lambda\) serves as a weighting factor to balance reconstruction and perception losses. By minimizing \(L_{r}\), the system will enhance the accuracy of the transmitted and reconstructed messages, focusing on bit-level precision. Minimizing \(L_{p}\) targets the reduction of perception-related errors, focusing on semantic-level understanding. Through careful adjustment of hyperparameter \(\lambda\), the system achieves an optimal balance and ensures both high fidelity in message reconstruction and semantic accuracy.
\iffalse 
\begin{algorithm2e}
\SetAlgoLined
\KwIn{The raw data $ X_i$  (LiDAR point clouds)}%输入参数
\KwOut{The detection output $\hat{Y}_i$}%输出
% \KwResult{Write here the result}
 \textbf{Step 1}: Train the semantic encoder and decoder at the first transmission. 
 \\
 \While{ not converge}{
   \For{$X_i$ in batch samples}{
         $F_i=\Phi(X_i)$\;
         $M_i^{(1)}={F_i}\odot{C_i}^{(1)}, C_i^{(1)}=P(F_i)$\;
         $T_i^{(1)}=\Psi_s^{(1)}(M_i^{(1)})$\;
         $\hat{M}_i^{(1)}=\Psi_d^{(1)}(\hat{T}_i^{(1)})$\;
   }
   Compute $L_{total}$\;
   Update the network $\Psi_s^{(1)}(\cdot)$,$\Psi_d^{(1)}(\cdot)$ using SGD;
   }
\textbf{Step 2}:  Train the semantic encoder and decoder at the $t$th ($t>2$) transmission. 
 \\
 \While{ not converge}{
   \For{$X_i$ in batch samples}{
         $F_i=\Phi(X_i)$\;
         $M_i^{(t)}={F_i}\odot{C_i}^{(t)}, C_i^{(t)}=P(F_i)$\;
         $T_i^{(t)}=\Psi_s^{(2)}(M_i^{(t)})$\;
         $\hat{M}_i^{(t)}=\Psi_d^{(2)}(\hat{T}_i^{(t)})$\;
   }
   Compute $L_{total}$\;
   Update the network $\Psi_s^{(2)}(\cdot)$,$\Psi_d^{(2)}(\cdot)$ using SGD;
   }

 \caption{Training Algorithm of SimHARQ-II}
\end{algorithm2e}
\fi

As demonstrated in Algorithm 1, the training process for the proposed model without HARQ unfolds in two distinct phases. Initially, we train the semantic encoder and decoder until convergence, generating \(\hat{M}_i^{(1)}\) and computing reconstruction loss. After calculating gradients, we update \(\Psi_s^{(1)}(\cdot)\) and \(\Psi_d^{(1)}(\cdot)\) using SGD. Next, we integrate detection output \(\hat{Y}_i\) for further loss computation. This step-by-step training avoids slow or divergent convergence and enhances semantic performance, ensuring the network not only reconstructs features accurately but also excels semantically.

\subsubsection{Training algorithm with HARQ}
Given that SimHARQ-I is designed for repeated feature transmissions, its semantic encoder \(\Psi_s^{(1)}(\cdot)\) and decoder \(\Psi_d^{(1)}(\cdot)\), trained as previously described, are same for both initial transmission and retransmissions. Then, we delineate a dual-phase training strategy for SimHARQ-II, aligning with this transmission methodology. The initial phase involves training \(\Psi_s^{(1)}(\cdot)\) and \(\Psi_d^{(1)}(\cdot)\) in a manner similar to the aforementioned process, preparing them for the initial transmission. The subsequent phase targets training \(\Psi_s^{(2)}(\cdot)\) and \(\Psi_d^{(2)}(\cdot)\) under conditions of lower SNR, equipping them to restore semantic information from impaired signals. This two-step training approach enhances learning stability and ensures consistent semantic performance throughout the transmission process.

\section{Evaluation}

To assess the performance of our cooperative perception model augmented with HARQ and an importance map, we conduct evaluation using the architecture depicted in Fig. \ref{fig:network}. The effectiveness of models is compared using the OPV2V dataset \cite{xu2022opv2v}, a detailed V2V cooperative perception dataset created through simulation on OpenCDA \cite{xu2022opv2v} and Carla platforms \cite{dosovitskiy2017carla}. This dataset includes 12,000 frames of 3D point clouds and RGB images, annotated with 230,000 3D bounding boxes within a \(40\hspace{2pt} m \times 40 \hspace{2pt} m\) perception zone. Our model utilizes the PointPillar framework for Lidar-based 3D object detection, which efficiently converts point cloud data into a bird’s eye view format. For back-propagation optimization, the Adam optimizer is employed to refine the perception performance.

We assess perception performance using average precision (AP) at intersection-over-union (IoU) thresholds of 0.50 and 0.70. AP, an established metric for object detection evaluation, measures algorithmic accuracy by averaging precision across varying recall levels. Higher AP values signify better object detection reliability. AP$@0.5$ typically exceeds AP$@0.7$, reflecting the stricter overlap criterion at the higher threshold.

% The channel SNR can be mathematically expressed as
% \begin{equation}
%     \text{SNR} = 10\log_{10} \frac{P}{{\sigma}^2},
% \end{equation}
% where \( P \) is the average power of the encoded signal and \( {\sigma}^2 \) is the average power of the noise in the channel. In our proposed scheme, \( P \) specifically represents the normalized power level of the channel input signal, which is usually set to 1. 

\subsection{Comparison with  Separate Coding Schemes}
First, our proposed method is compared with a conventional separate source and channel coding scheme without employing HARQ. The baseline scheme utilizes uniform quantization for processing feature \(M_i\) obtained from the backbone network. For source coding, 8-bit quantization is adopted. It is important to note that the perception performance remains unaffected by the 8-bit quantization by carefully selecting an appropriate quantization step size and zero point. Additionally, for channel coding, we consider a 1/2-rate Low-Density Parity-Check (LDPC) code with a code length of 5,376. Following the channel coding, modulation is performed using either 16-QAM or 256-QAM. This comparison is designed to offer insights into the performance enhancements provided by our method, especially in scenarios where HARQ is not utilized.
\begin{table}[ht] 
\centering
\caption{Simulation parameters for OFDM channel.}
\begin{tabular}{|c|c|} \hline 
 Parameters & Values  \\ \hline
  Number of subcarriers & 2048\\ \hline
Number of OFDM symbols & 14\\ \hline
Subcarrier spacing & 15 kHz \\ \hline
Carrier frequency &  2.8 GHz\\ \hline
Retransmission interval &  2 ms\\ \hline
\end{tabular}
\label{tab:ofdm}
\end{table}

Meanwhile, various channel coding schemes, when combined with different modulation techniques at a specific $\text{CR}^{(1)}$, may require different numbers of channel uses to transmit the same input. To maintain a fair comparison, we calibrate the CR to equalize the channel uses across different modulation and coding schemes (MCS). The volume of the raw data is denoted by \(D_r\) and the channel uses for the baseline schemes are calculated as
\begin{equation}
    \text{Channel uses} = \frac{D_r \times \text{CR}^{(1)} \times 8}{\log_2{M_c} \times R_c},
\end{equation}
where \(R_c\) is the coding rate and \(M_c\) is the QAM order. Accordingly, $\text{CR}^{(1)}$ of $1.25 \times 10 ^{-3}$ and $2.5 \times 10 ^{-3}$ can be set for 1/2-rate LDPC with 16QAM and 256QAM, respectively, to ensure the same number of channel uses as our proposed methods with a CR of $5 \times 10 ^{-3}$. The scheme with a CR of $2.5 \times 10 ^{-3}$, encoding more information bits, consequently requires a higher order of modulation to accommodate the additional data.

We train and test our proposed system over a time-varying multipath fading channel, which is modeled based on QuaDRiGa \cite{jaeckel2014quadriga}. QuaDRiGa can support the simulation of moving transmitters and receivers (e.g. for car-to-car or device-to-device communication) through setting the tracks. In simulation, we assume that the speeds  of vehicles range from 40 to 100 kilometers per hour (km/h) and the distances between vehicles vary from 28 to 70 meters (m).

% \begin{figure}[htbp]
%     \centering
%     \includegraphics[width=0.5\textwidth]{numpath-ap0.7.pdf}
%     \caption{Robustness test of our proposed method for the number of paths.}
%     \label{fig:path}
% \end{figure}
\begin{figure}[htp]
    \centering
    \includegraphics[width=0.8\textwidth]{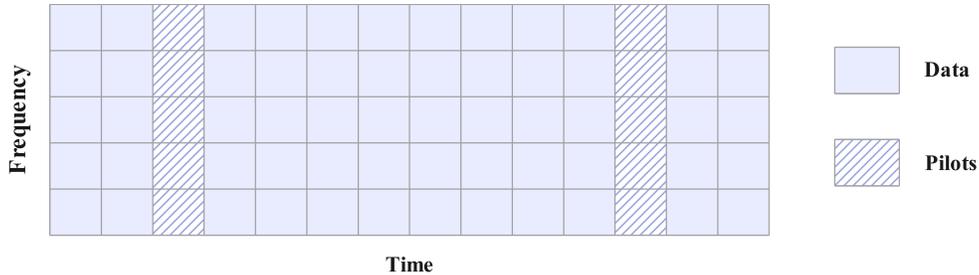}
    \caption{Kronecker-based pilot for channel estimation of a time-varying OFDM channel.}
    \label{fig:pilot}
\end{figure}

As detailed in Table \ref{tab:ofdm}, our configuration for the OFDM system includes setting the number of subcarriers \(L_{\text{fft}}\) to 2,048, with a subcarrier spacing of 15 kHz and a carrier frequency of 2.8 GHz. For the transmission of a LiDAR point frame, we allocate 12 symbols for information and 2 symbols for pilots, enabling the entire frame to be conveyed across 14 OFDM symbols. In scenarios where the acknowledgment signal \(\text{ACK}\) is 0, indicating a need for retransmission, the transmitter is programmed to resend the feature after a 1 ms delay. The structure for the Kronecker-based pilot channel estimation, suited for the proposed time-varying OFDM channel, is visualized in Fig. \ref{fig:pilot}. Within this structure, the third and twelfth symbols are specifically allocated for the transmission of pilot signals. This arrangement allows for the entire channel, encompassing 14 OFDM symbols, to be accurately estimated via linear interpolation, effectively capturing the time-varying channel state information (CSI). 
\begin{figure*}[htbp]
    \begin{minipage}[t]{\linewidth}
        \centering
        \includegraphics[width=0.55\textwidth]{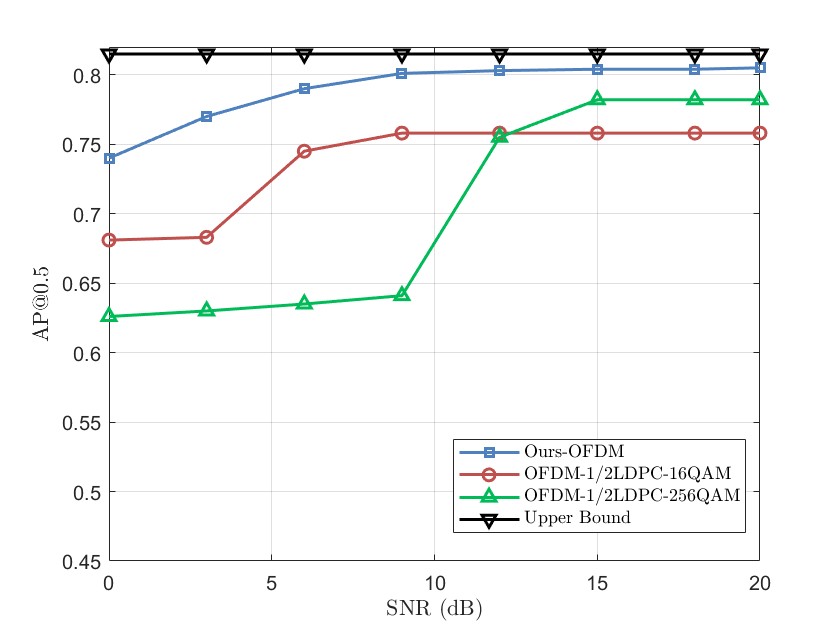}
        \centerline{(a)}
    \end{minipage}%
    \vfill
    \begin{minipage}[t]{\linewidth}
        \centering
        \includegraphics[width=0.55\textwidth]{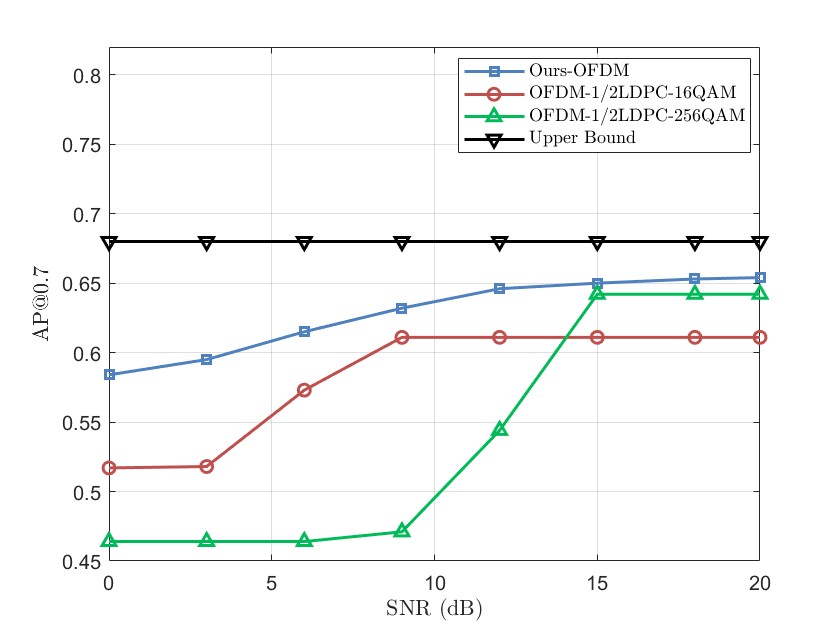}
        \centerline{(b)}
    \end{minipage}
    \caption{Performance of our proposed method compared with baseline schemes at different SNRs over a time-varying multipath fading channel. The baseline schemes and our proposed method exploit the same channel uses for fairness. (a) Average precision performance  at the IoU threshold 0.5; (b) Average precision performance  at the IoU threshold 0.7.}
    \label{fig:base}
\end{figure*}

Fig. \ref{fig:base} illustrates the performance of the proposed method compared with baseline schemes at different SNRs over a time-varying OFDM channel. The upper bound can be obtained assuming perfect communication with  $\text{CR}^{(1)}=1$. In Fig. \ref{fig:base}(a), our proposed method outperforms the separate coding schemes with the same channel uses in terms of AP$@0.5$ since our proposed method is designed and trained with JSCC, which can jointly optimize the whole system to preserve more semantic information given the same communication resources used. Meanwhile, the proposed method exhibits smoother performance in low SNR regimes without the cliff effect, which is commonly observed in the traditional communication systems. Hence, our proposed method can prevent catastrophic perception performance loss in the low SNR regimes, which is critical to autonomous driving. Moreover, our method significantly outperforms the traditional schemes and approaches the performance upper bound constrained by the cooperative perception module at SNR = 9 dB. On the one hand, traditional coding combined with low-order modulation achieves better performance in the low SNR regimes and reaches its maximum performance at SNR = 9 dB. On the other hand, despite performing worse than low-order modulation in the low SNR regimes, high-order modulation reaches its maximum performance at SNR = 15 dB and surpasses the performance of low-order modulation. This is because high-order modulation allows the transmission of more semantic information, especially in high SNR regimes.

In Fig. \ref{fig:base}(b), the performance of our approach is compared with that of the conventional methodologies in terms of AP$@0.7$. Despite outperforming traditional methods, the AP$@0.7$ performance is worse than AP$@0.5$. This degradation can be attributed to the stricter evaluation at the higher IoU threshold, making AP$@0.7$ more susceptible to noise and requiring more precise information. In our proposed methods, AP$@0.5$ nearly reaches the upper limit at an SNR of 9 dB, whereas AP$@0.7$ gets close to this upper limit at an SNR of 15 dB. Given the need for more precise information at AP$@0.7$, the performance of our method is slightly below that of perfect communication scenarios. This highlights the importance of accurate and reliable information transmission to maintain high performance at stricter IoU thresholds, especially in applications where accurate detection is crucial.

\subsection{Performance of SimHARQ-I}

In this subsection, we investigate the performance of our proposed SimHARQ-I method. As a baseline, we maintain the traditional communication channel and source coding parameters consistent with the previous section. For the traditional communication methods, retransmission is facilitated using HARQ-I, which includes the utilization of a 24-bit CRC for error detection. The maximum number of transmissions is capped at three ($B=3$).

\begin{figure*}[htbp]
    \begin{minipage}[t]{\linewidth}
        \centering
        \includegraphics[width=0.55\textwidth]{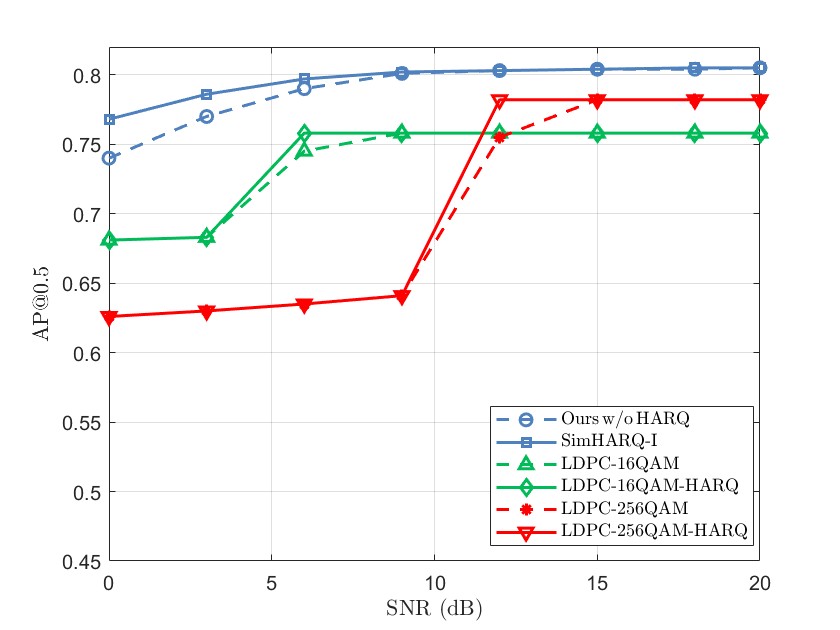}
        \centerline{(a)}
    \end{minipage}%
    \vfill
    \begin{minipage}[t]{\linewidth}
        \centering
        \includegraphics[width=0.55\textwidth]{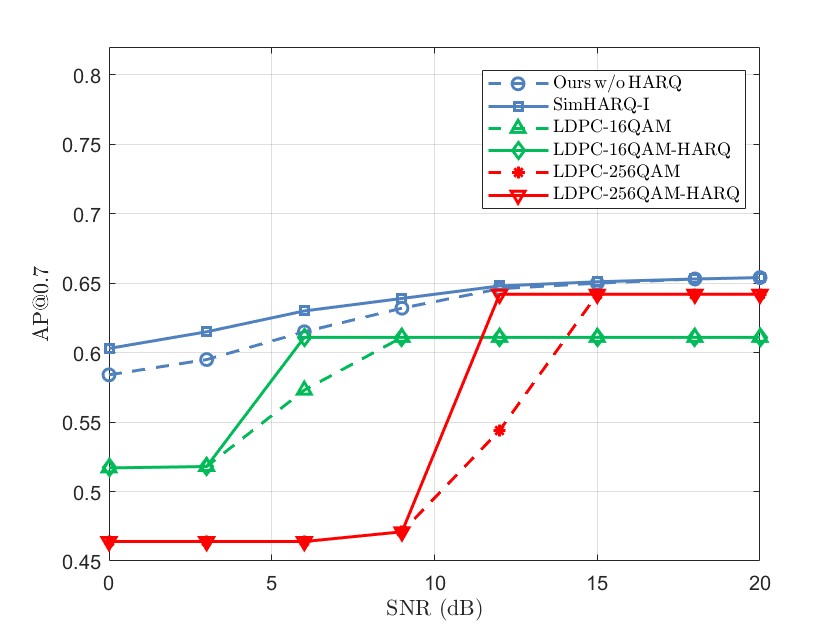}
        \centerline{(b)}
    \end{minipage}
    \caption{Performance of the proposed SimHARQ-I compared with the conventional HARQ-I. (a) Average precision performance  at the IoU threshold 0.5; (b) Average precision performance  at the IoU threshold 0.7.}
    \label{fig:HARQ-I}
\end{figure*}

We first compare the performance of our proposed SimHARQ-I with that of the conventional system combined with HARQ-I with error detection threshold $\beta=0.72$. As illustrated in Fig.~\ref{fig:HARQ-I}, the incorporation of HARQ with our method (`SimHARQ-I') demonstrates a noticeable improvement in performance, indicating that HARQ is effective in enhancing the resilience and reliability of the communication link, especially in the low SNR regimes. Fig. \ref{fig:HARQ-I}(b) delineates a distinct plateau in performance enhancement beyond an SNR of 15 dB for SimHARQ-I, suggesting the SimHARQ-I is not particularly useful under conditions of high SNR regimes. In high SNR scenarios,  most of the first transmissions are semantically accurate due to the significant reduction of semantic errors, resulting in minimal gain from HARQ. For all baseline schemes, significant semantic errors from channel noise render HARQ-I ineffective in the low SNR scenarios. In the moderate SNR scenarios, there is an approximate 3 dB gain while in high SNR scenarios, there is no gain. Across all SNRs, SimHARQ-I maintains a 
better performance than the conventional methods, thereby demonstrating not only the efficacy of the proposed JSCC methods but also its enhanced robustness combined with HARQ protocols.
\begin{figure}[htbp]
    \centering
    \includegraphics[width=0.55\textwidth]{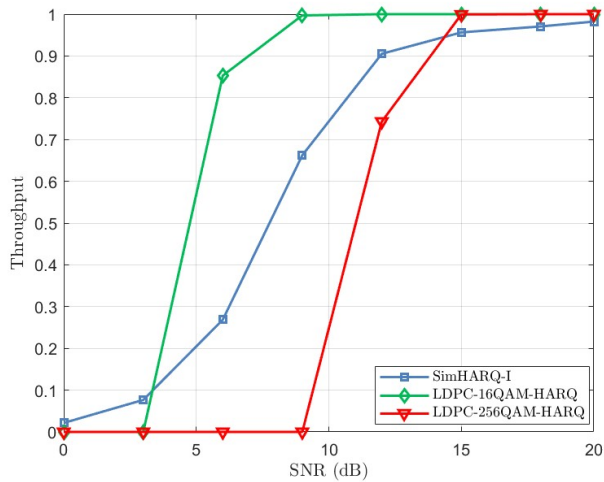}
    \caption{Throughput of our proposed SimHARQ-I compared with the conventional HARQ-I.}
    \label{fig:throughput}
\end{figure}

Fig. \ref{fig:throughput} illustrates the throughput performance of various HARQ schemes across different SNRs. In this paper, throughput is defined as the ratio of the number of successful transmissions to the total number of transmissions. It shows that the lower-order modulation scheme, LDPC-16QAM-HARQ, achieves a throughput of 1 at a lower SNR compared to higher-order schemes, such as LDPC-256QAM-HARQ. This efficiency is due to the superior error correction capabilities of lower-order modulations, enabling nearly error-free transmission at SNRs as low as 9 dB. Although the throughput of SimHARQ-I is lower than that of lower-order modulation schemes at the low SNRs, our scheme significantly surpasses traditional schemes in AP performance. Moreover, compared to the traditional higher-order modulation schemes, our scheme exhibits outstanding throughput performance under low SNR conditions, surpassing these traditional approaches. This highlights that our scheme, while maintaining communication efficiency, can also ensure high data transmission quality under complex channel conditions, especially in low SNR environments, indicative of the effective employment of the semantic error correction strategy, SimCRC.

\begin{figure*}[htbp]
    \begin{minipage}[t]{\linewidth}
        \centering
        \includegraphics[width=0.54\textwidth]{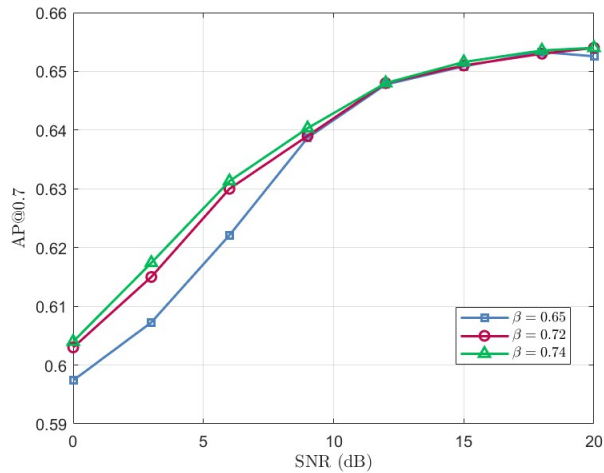}
        \centerline{(a)}
    \end{minipage}%
    \vfill
    \begin{minipage}[t]{\linewidth}
        \centering
        \includegraphics[width=0.54\textwidth]{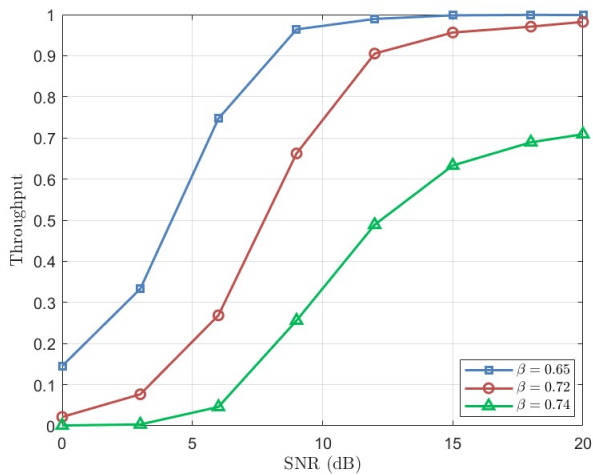}
        \centerline{(b)}
    \end{minipage}
    \caption{Robustness test of our proposed method for the error detection threshold $\beta$. (a) Average precision performance  at the IoU threshold 0.7; (b) Throughput.}
    \label{fig:beta-I}
\end{figure*}

Fig. \ref{fig:beta-I} illustrates the impact of parameter $\beta$ on system performance over SNRs in terms of AP@$0.7$ and throughput. As Fig. \ref{fig:beta-I}(a) illustrates, there is a noticeable improvement in performance as $\beta$ increases. Specifically, the curve for $\beta=0.72$ lies significantly above the curve for $\beta=0.65$, indicating better performance in all SNR regimes. However, as depicted in Fig. \ref{fig:beta-I}(b), when $\beta$ is set too high, for example at $\beta=0.74$, there is only a marginal gain in performance, but the throughput declines sharply, especially noticeable beyond 10 dB SNR. This decline in throughput with a higher $\beta$ can be attributed to a stricter error detection regime, which prompts more frequent retransmissions to ensure accuracy. On the contrary, a smaller $\beta$, such as $\beta=0.65$, leads to a much higher throughput. The system tends to accept more features as correct, minimizing the number of retransmissions. Therefore, to optimize the system, we should choose an appropriate $\beta$ to achieve a balance between   performance and throughput.

\subsection{Performance of SimHARQ-II}

In this section, we will compare the conventional HARQ Type II scheme, which utilizes LDPC coding, and our proposed SimHARQ-II. The LDPC codes are configured with redundancy versions (RVs) of $[0,2,1]$, adhering to the specifications provided by the 3GPP standards \cite{3gpp38212}, \cite{3gpp38214}. To strike an optimal balance between latency and reliability, the maximum number of transmissions is capped at three ($B=3$). The initial transmission employs LDPC-encoded information with an RV of 0, complemented by a 24-bit CRC for error detection. Should an error be identified, subsequent transmissions utilize information encoded with RVs of 2 and 1, respectively, in sequence, until a successful transmission is achieved. It is crucial to highlight that the coding rate for all transmissions is maintained at 1/2, with a codeword length of 5,376.
\begin{figure*}[htbp]
    \begin{minipage}[t]{\linewidth}
        \centering
        \includegraphics[width=0.55\textwidth]{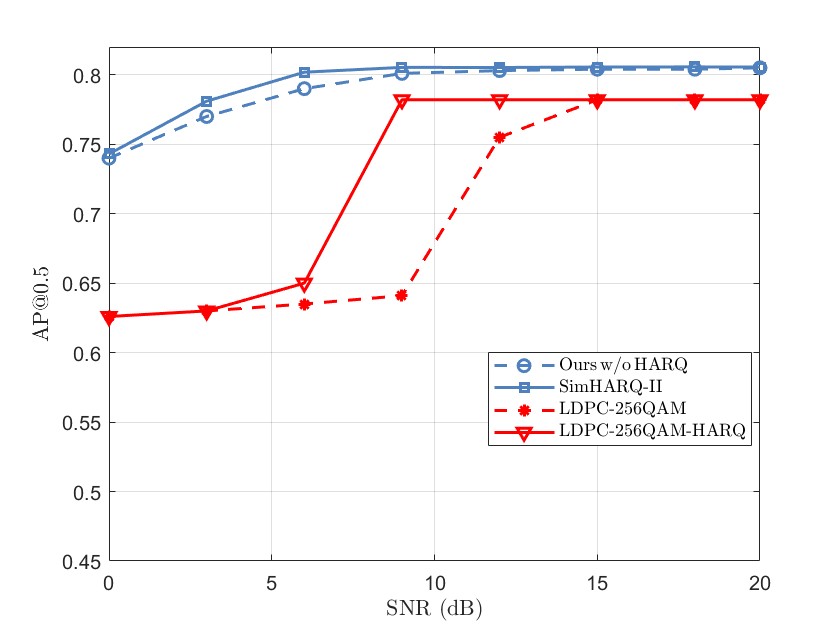}
        \centerline{(a)}
    \end{minipage}%
    \vfill
    \begin{minipage}[t]{\linewidth}
        \centering
        \includegraphics[width=0.55\textwidth]{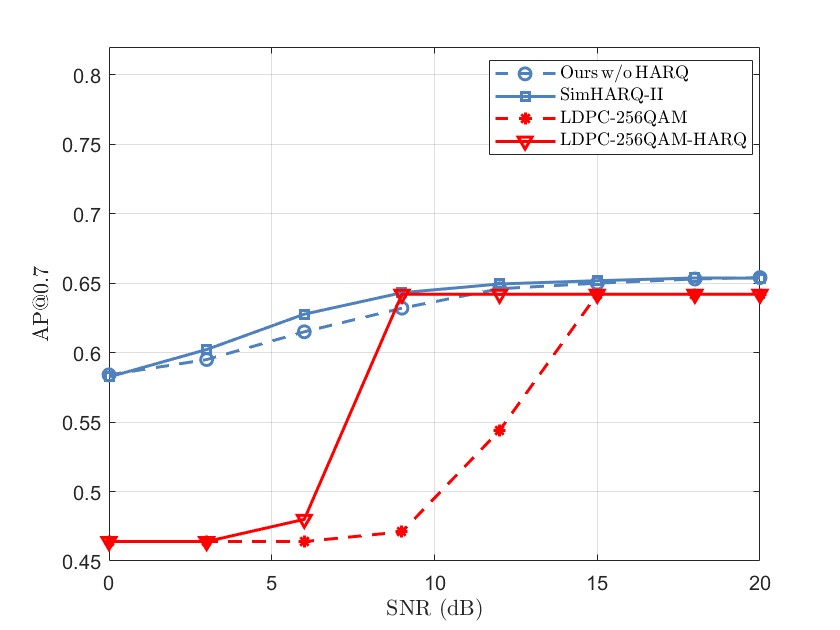}
        \centerline{(b)}
    \end{minipage}
    \caption{Performance of the proposed SimHARQ-II compared with the conventional HARQ-II. (a) Average precision performance  at the IoU threshold 0.5; (b) Average precision performance  at the IoU threshold 0.7.}
    \label{fig:HARQ-II}
\end{figure*}

Fig. \ref{fig:HARQ-II} illustrates the performance of traditional HARQ-II and our proposed SimHARQ-II with respect to AP with error detection threshold $\beta=0.72$. It is evident that retransmission has a significant positive impact on both traditional communication schemes and our method. For the traditional communication system, incremental HARQ shows a substantial gain, with an improvement of approximately 6 dB observable in terms of AP@$0.5$ and AP@$0.7$. For instance, the performance at 9 dB with HARQ-II nearly matches the performance at 15 dB without HARQ, highlighting the efficiency of the retransmission process. Similarly, Fig. \ref{fig:HARQ-II}(a) for AP@0.5 also indicates a remarkable gain due to the SimHARQ-II implementation in our method. The gain is around 6 dB, analogous to the improvement in the traditional scheme. This is illustrated by the fact that the performance at 6 dB with HARQ is comparably close to the performance at 12 dB, underscoring the effectiveness of SimHARQ-II in boosting system reliability even at lower SNR levels. When considering the AP@0.7 depicted in Fig. \ref{fig:HARQ-II}(b), the benefit of SimHARQ-II appears to be slightly less pronounced, yet still significant. Therefore, while SimHARQ-II contributes to enhancing performance, the extent of improvement may vary across different AP thresholds. It suggests that for higher AP thresholds, the system may require additional retransmissions to achieve similar gains. Furthermore, it is important to note that our proposed method consistently outperforms the traditional communication system across the entire SNR regimes. This superior performance suggests that our method benefits significantly not only from the utilization of HARQ but also from the employment of JSCC. 
\begin{figure}[htbp]
    \centering
    \includegraphics[width=0.5\textwidth]{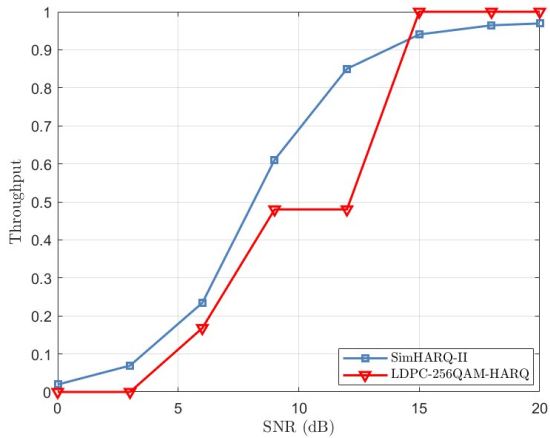}
    \caption{Throughput of our proposed SimHARQ-II compared with the conventional HARQ-II.}
    \label{fig:throughput-II}
\end{figure}

Fig. \ref{fig:throughput-II} compares the throughput of the traditional HARQ-II and our proposed SimHARQ-II across a range of SNRs. From the figure, our proposed method outperforms the traditional approach, particularly in the low SNR scenarios. This indicates that our method is more efficient in terms of communication resource utilization, as it achieves higher throughput in lower SNR regimes. For instance, at an SNR of 9 dB, both methods achieve similar performance in terms of AP@$0.7$, however, our proposed method demonstrates superior throughput, suggesting enhanced efficiency. On the other hand, as the SNR increases, the traditional method slightly outperforms our proposed method in terms of throughput, yet the perception performance suffers according to Fig. 13(b). The overall assessment indicates that our method conserves communication resources more effectively.

\section{Conclusion}
In this paper, we have introduced a novel JSCC framework combined with HARQ specifically designed for the transmission and intermediate fusion of LiDAR point clouds over wireless channels. This architecture features a semantic encoder that directly translates the input LiDAR point clouds into channel inputs, employing CNNs for both encoding and decoding functions. These CNNs are trained in the end-to-end manner, focusing on minimizing both the cooperative perception and reconstruction losses to optimize performance. To address the critical need for reliable transmission, particularly in environments with low SNR, we have developed a novel semantic error detection method, which  significantly boosts transmission reliability when paired with our semantic communication framework and HARQ. Our simulation results clearly demonstrate the superiority of our model over traditional separate source-channel coding techniques in terms of perception performance, affirming the robustness of our system with or without HARQ implementation. Moreover, our proposed HARQ strategies showcase remarkable efficiency in throughput, outperforming conventional coding methods and highlighting the overall effectiveness of our proposed framework.


%\bibliographystyle{elsarticle-num}
%\bibliographystyle{cas-model2-names}
%\bibliographystyle{ieeetr}
\small
\bibliographystyle{IEEEtran}
%\bibliographystyle{unsrt}

\bibliography{journal_dc.bib}

\end{document}